\documentclass{article}

\usepackage{arxiv}

\usepackage[utf8]{inputenc} 
\usepackage[T1]{fontenc}    
\usepackage{hyperref}       
\usepackage{url}            
\usepackage{booktabs}       
\usepackage{amsfonts}       
\usepackage{nicefrac}       
\usepackage{microtype}      
\usepackage{graphicx}
\usepackage{lipsum}

\title{Dielectron production in proton--proton collisions at $\sqrt{s} = 7$\,TeV with ALICE}

\newcommand{\DCAee}{$\rm DCA_{ee}$}

\author{
  H. Sebastian Scheid\\
  for the ALICE Collaboration\\
  \texttt{s.scheid@cern.ch} \\
}

\begin{document}
\maketitle

\begin{abstract}
The ALICE Collaboration measured dielectron production as a function of the invariant mass ($m_{\rm ee}$), the pair transverse momentum ($p_{\rm T,ee}$) and the pair distance of closest approach (\DCAee) in pp collisions at $\sqrt{s} = 7$\,TeV. Prompt and non-prompt dielectron sources can be separated with the \DCAee, which will give the opportunity in heavy-ion collisions to identify thermal radiation from the medium in the intermediate-mass range dominated by contributions from open-charm and beauty hadron decays. The charm and beauty total cross sections are extracted from the data by fitting the spectra with two different MC generators, i.e. PYTHIA a leading order event generator and POWHEG a next-to-leading order event generator. Significant model dependences are observed, reflecting the sensitivity of this measurement to the heavy-flavour production mechanisms.\footnote{Presented at Hot Quarks 2018 - Workshop for young scientists on the physics of ultrarelativistic nucleus-nucleus collisions, Texel, The Netherlands, September 7-14 2018}
\end{abstract}


\section{Introduction}
The ALICE experiment at the LHC is dedicated to the study of strongly interacting matter under extreme conditions, i.e. high temperature, which can be reached in heavy-ion collisions. In such collisions, the formation of a Quark-Gluon Plasma (QGP) is expected. Dielectrons are produced at all stages of the collision and therefore carry information about the whole evolution of the system. Since they do not interact strongly with the medium, they are a prime probe to study the properties of the QGP. Dielectrons stem from decays of pseudoscalar and vector mesons, from semi-leptonic decays of correlated open-charm and open-beauty hadrons and from internal conversions of direct photons. In heavy-ion collisions, additional sources are expected, i.e. thermal radiation from the QGP and hadron gas. The medium introduces modifications of the vector meson properties, in particular the short-lived $\rho$, related to chiral symmetry restoration. In addition, the initial conditions of the collisions are expected to change compared to elementary collisions due to modifications of the parton distribution functions in nuclei. The latter can be studied in proton-lead (p--Pb) collisions, whereas pp collisions provide an important vacuum baseline. It is crucial to first understand the dielectron production in vacuum to single out the signal characteristics of the QGP. Moreover, proton-proton (pp) collisions can also be used to study the heavy-flavour (HF) and direct photon production.

In the following, the steps of the data analysis are explained and the first measurements of the dielectron production in pp collisions at $\sqrt{s} = 7$\,TeV are presented and discussed~\cite{ref-ee}.

\section{Data analysis and results}


The analysis is performed with pp data taken during the first data-taking period of the LHC in 2010 with the ALICE detector. The integrated luminosity of the data sample is $L_{\rm int} = 6.0\pm0.2$\,nb$^{-1}$.
After identifying electrons in the ALICE detector it is not a priori clear which electrons belong to the same pair. We follow a statistical approach to obtain the final spectrum. The electrons and positrons are combined to an opposite-sign spectrum (OS), which includes not only the signal but also background, that can be purely combinatorial or have some residual correlation from jets or cross pairs from double Dalitz decays. This background is estimated by the same-sign spectrum (SS). Residual acceptance differences for OS and SS pairs are estimated with mixed events and taken into account during the subtraction of the background. Finally, the background-subtracted spectrum is corrected for tracking and particle identification inefficiencies within the ALICE acceptance ($p_{\rm T,e} > 0.2$\,GeV/$c$, $ \eta_{\rm e}<0.8 $).

In Fig. 1 the measured dielectron cross section as a function of $m_{\rm ee}$ is compared to a so-called hadronic cocktail, which includes all known sources of dielectron production from hadron decays and uses parameterisations of measured spectra as input when available. Where no measurements are available $m_{\rm T}$-scaling~\cite{ref-mt-scaling} is applied. The HF contributions are simulated using the Perugia2011 tune of PYTHIA~6~\cite{ref-pythia,ref-pythia2011}. The resulting dielectron pairs from the hadron decays are then filtered through the ALICE acceptance.

\begin{figure}
\centering
\includegraphics[scale=0.355]{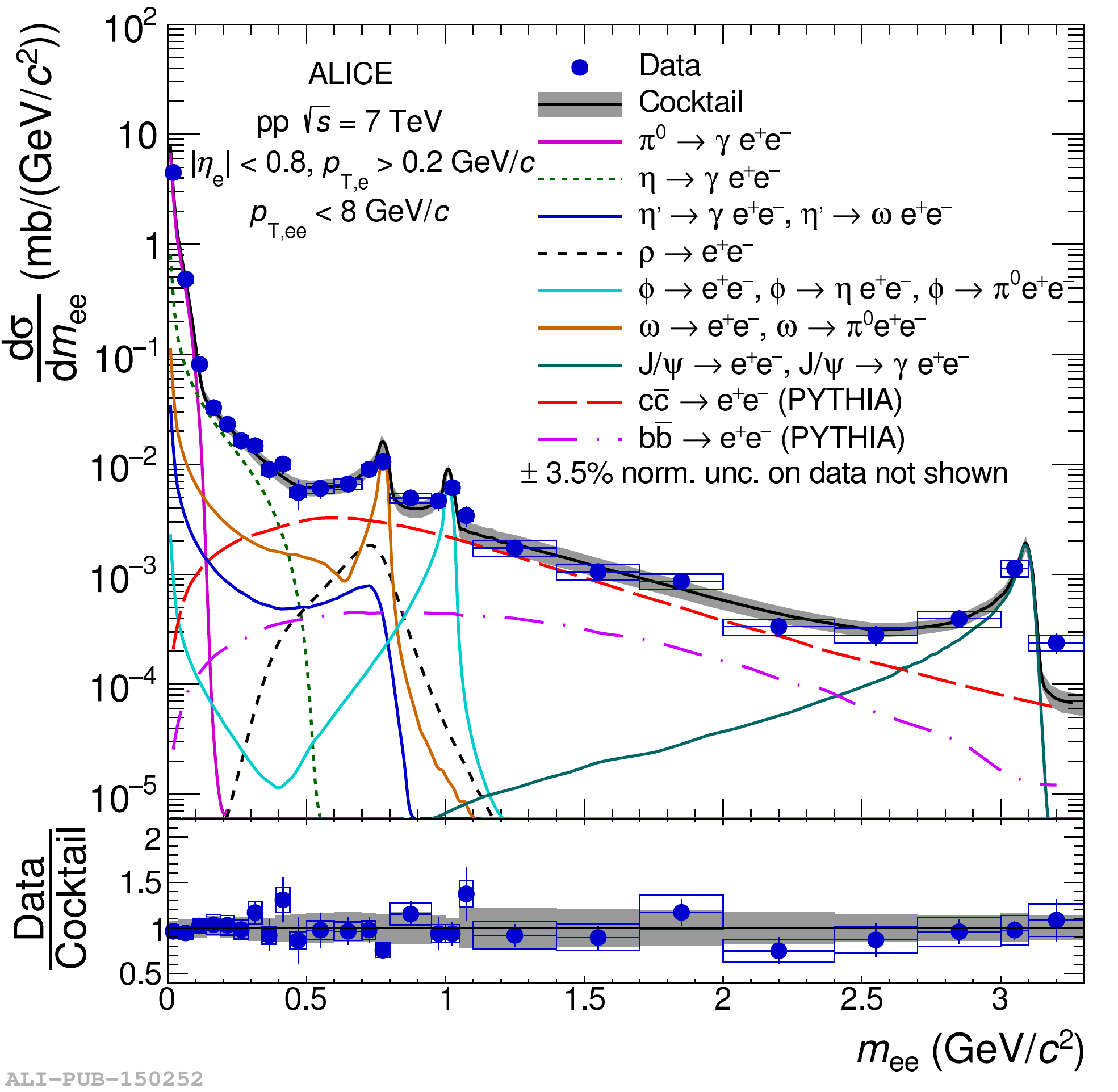}
\caption{Dielectron cross section as a function of $m_{\rm ee}$ compared to a cocktail of known hadronic sources.}
\end{figure}

A good agreement is observed between the cocktail and the data. The charm contribution already dominates the spectrum for $m_{\rm ee} \geq 0.5$\,GeV/$c^{2}$. The very large heavy-flavour contribution makes the measurement of thermal radiation from the medium in heavy-ion collisions very challenging at LHC energies. To separate the heavy-flavour background from thermal radiation from the QGP in a future heavy-ion run in the intermediate-mass range (IMR, $\phi < m_{\rm ee} < J/\psi$), an additional variable, the pair-distance-of-closest-approach (\DCAee), is added to the traditional analysis as a function of $m_{\rm ee}$ and $p_{\rm T,ee}$. \DCAee~is defined as:
\begin{equation}
{\rm DCA_{ee}} = \sqrt{\frac{({\rm DCA_{{\it xy},1}}/\sigma_{xy{ \rm ,1}})^{2}+({\rm DCA_{{\it xy},2}}/\sigma_{xy,2})^{2}}{2}}
\end{equation}

Here ${\rm DCA}_{xy,i}$ is the closest distance between the reconstructed electron track and the primary collision vertex in the transverse plane. Its resolution is estimated from the covariance matrix of the track reconstruction and denoted as $\sigma_{xy,i}$. In the case of weak decays, the decay electron candidates do not point to the vertex which leads to a broader DCA distribution than for tracks from prompt decays.
This can be seen in Fig. 2 and Fig. 3, where the \DCAee~spectra are shown for two invariant mass regions. Fig. 2 shows the mass region between the $\pi^{0}$ and the $\phi$ mass. The light flavour template is taken from the $\pi^{0}$ shape, normalised to the expected contribution from all light flavour sources. Fig. 3 shows the mass region around the $J/\psi$ mass peak. In both mass regions we can see a clear peak which can be described by the expected prompt contributions, whereas the tail of the spectrum is described by the broader contributions from charm and beauty.
\begin{figure}
\centering
  \begin{minipage}{0.47\textwidth}
    \includegraphics[scale=0.35]{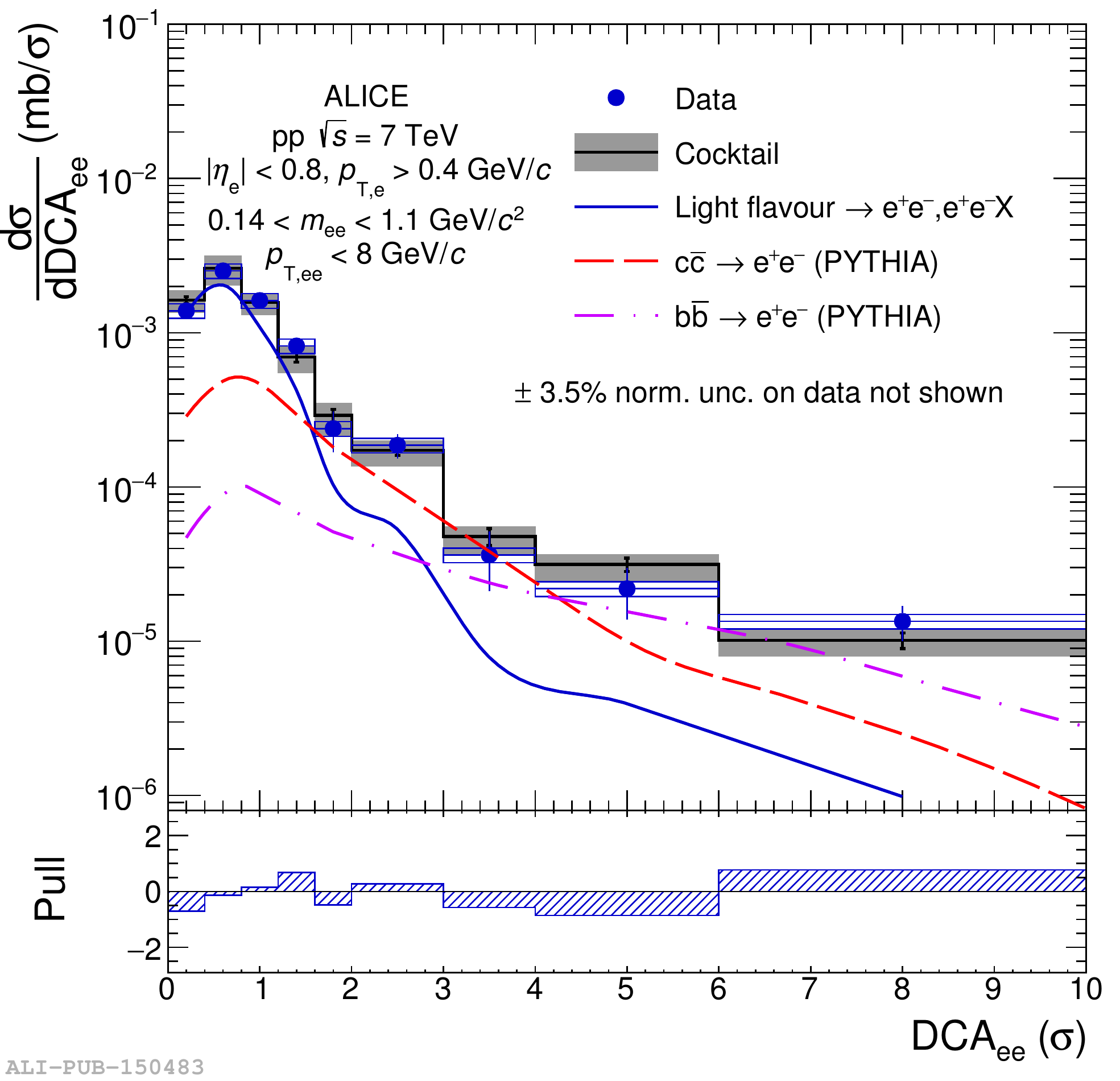}
    \caption{Dielectron spectrum as a function of \DCAee~for $0.14 < m_{\rm ee} < 1.1$\,GeV/$c^2$~\cite{ref-ee}.}
  \end{minipage}
  \begin{minipage}{0.47\textwidth}
    \includegraphics[scale=0.35]{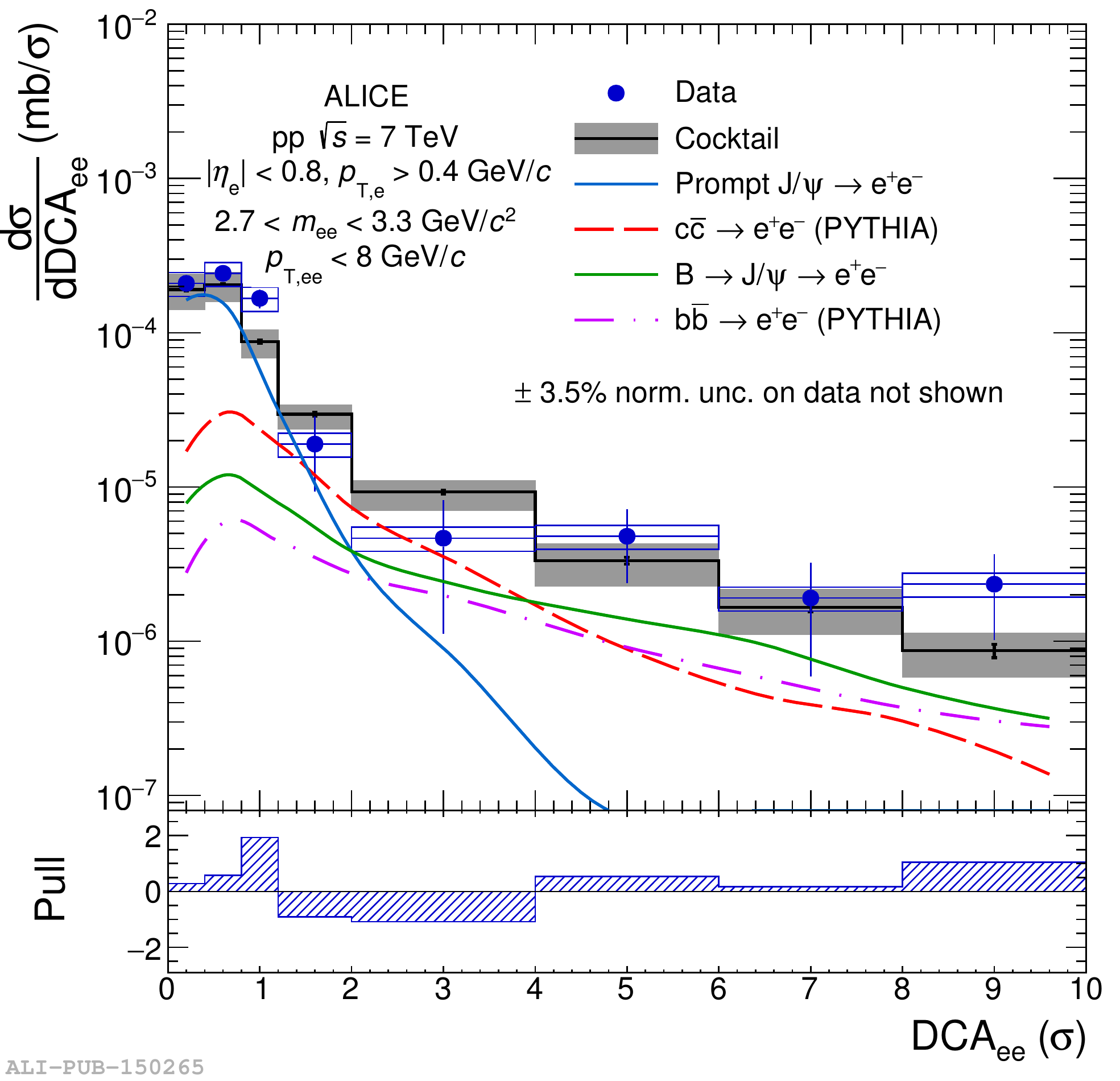}
    \caption{Dielectron spectrum as a function of \DCAee~for $2.7 < m_{\rm ee} < 3.3$\,GeV/$c^2$~\cite{ref-ee}.}
  \end{minipage}
\end{figure}
In Fig. 3 the $J/\psi$ from $B$-mesons  can be seen in addition to the open HF contributions.
In the so-called intermediate mass region, located between the $\phi$ and $J/\psi$ in the mass spectrum, the dominant contribution is from open HF.
The dielectron cross section as function of $p_{\rm T,ee}$ and \DCAee~is compared to a hadronic cocktail using PYTHIA 6 Perugia0~\cite{ref-pythia2011} to estimate the $\rm c\bar{c}$ and $\rm b\bar{b}$ contributions in the left and right panels of Fig. 4, respectively.
\begin{figure}
\centering
  \begin{minipage}{0.47\textwidth}
    \includegraphics[scale=0.35]{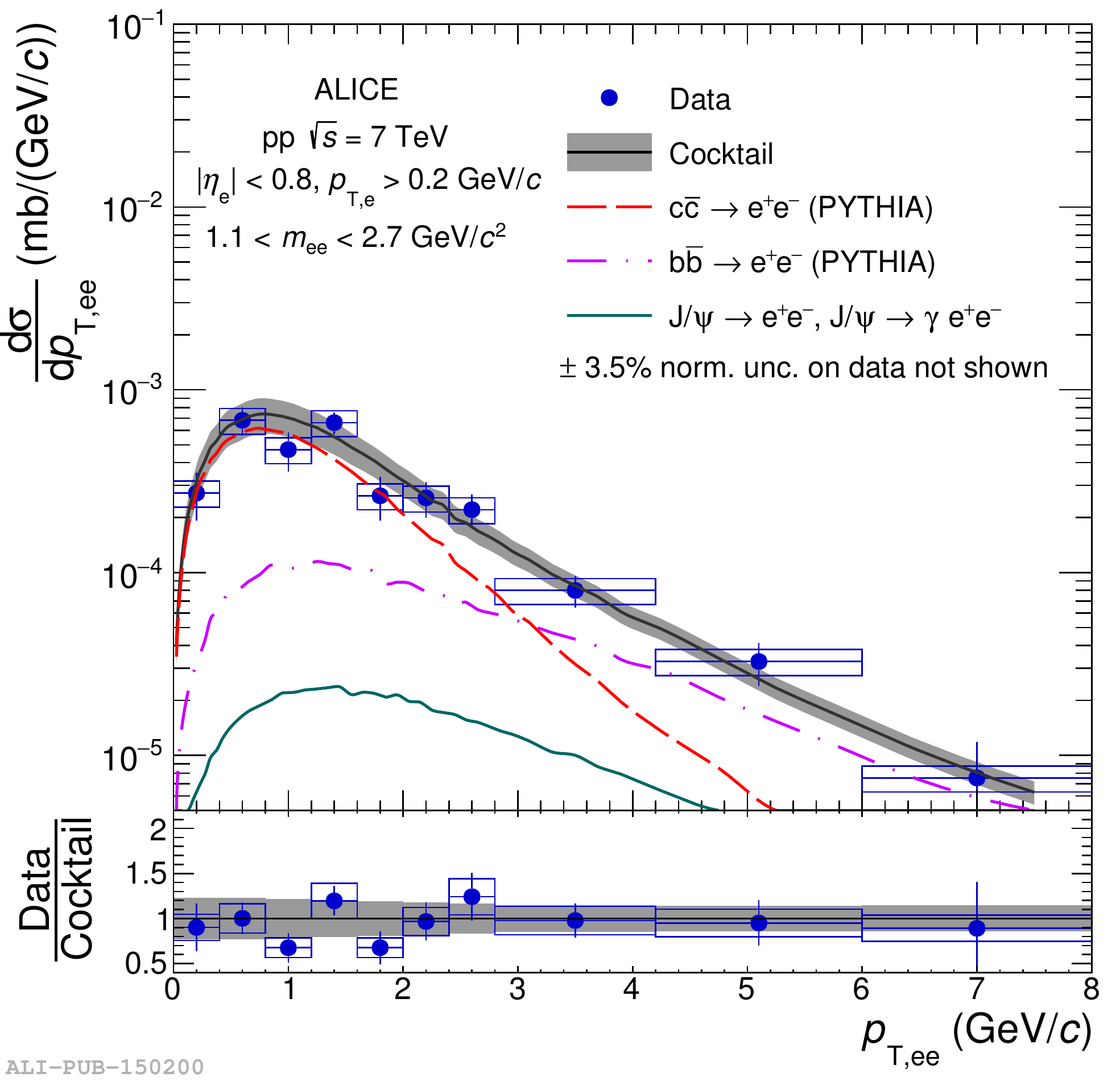}
  \end{minipage}
  \begin{minipage}{0.47\textwidth}
    \includegraphics[scale=0.35]{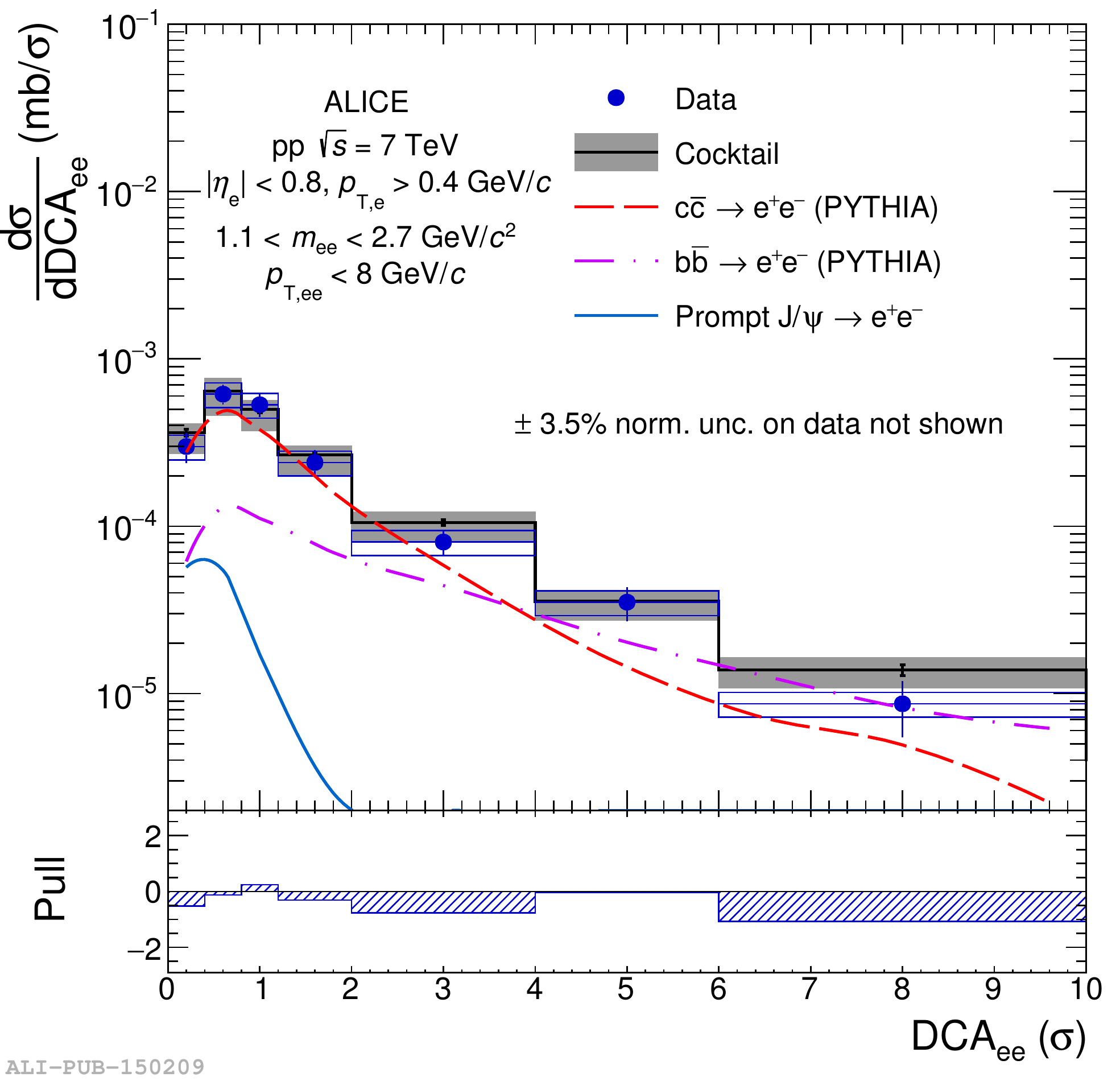}
  \end{minipage}
\caption{Dielectron cross section as a function of $p_{\rm T,ee}$ (left) and \DCAee~(right) in the IMR compared to a cocktail calculated with PYTHIA~6~\cite{ref-ee}.}
\end{figure}
The data are well described by the hadronic cocktail within the statistical and systematic uncertainties. The contribution from $\rm c\bar{c}$ dominates the dielectron yield at low $p_{\rm T,ee}$ and relatively small \DCAee, whereas the $\rm b\bar{b}$ becomes relevant at high $p_{\rm T,ee}$ and large \DCAee.
To investigate the processes of heavy-quark production we changed the generator from PYTHIA to POWHEG, switching from leading order in the HF quark generation to next-to-leading order. To quantify the differences the total $\rm c\bar{c}$ and $\rm b\bar{b}$ cross sections are extracted from the data by fitting the results two-dimensionally as a function of $p_{\rm T,ee}$ and $m_{\rm ee}$ and one-dimensionally as a function of \DCAee~in the IMR allowing the contributions of the two HF components to vary. The results are shown in the left and right panels of Fig. 5  for PYTHIA and POWHEG\cite{ref-powheg}, respectively.
Both fits give consistent results for a given MC event generator. The uncertainties are fully correlated between the cross sections extracted with PYTHIA and POWHEG. Significant model dependences are observed which reflect the different rapidity distribution of charm quarks and different $p_{\rm T,ee}$ spectra of the $\rm c\bar{c}$ and $\rm b\bar{b}$ contributions predicted by the two models.
\begin{figure}
\centering
  \begin{minipage}{0.47\textwidth}
    \includegraphics[trim={0, 0, 0, 1.5cm},clip,scale=0.357]{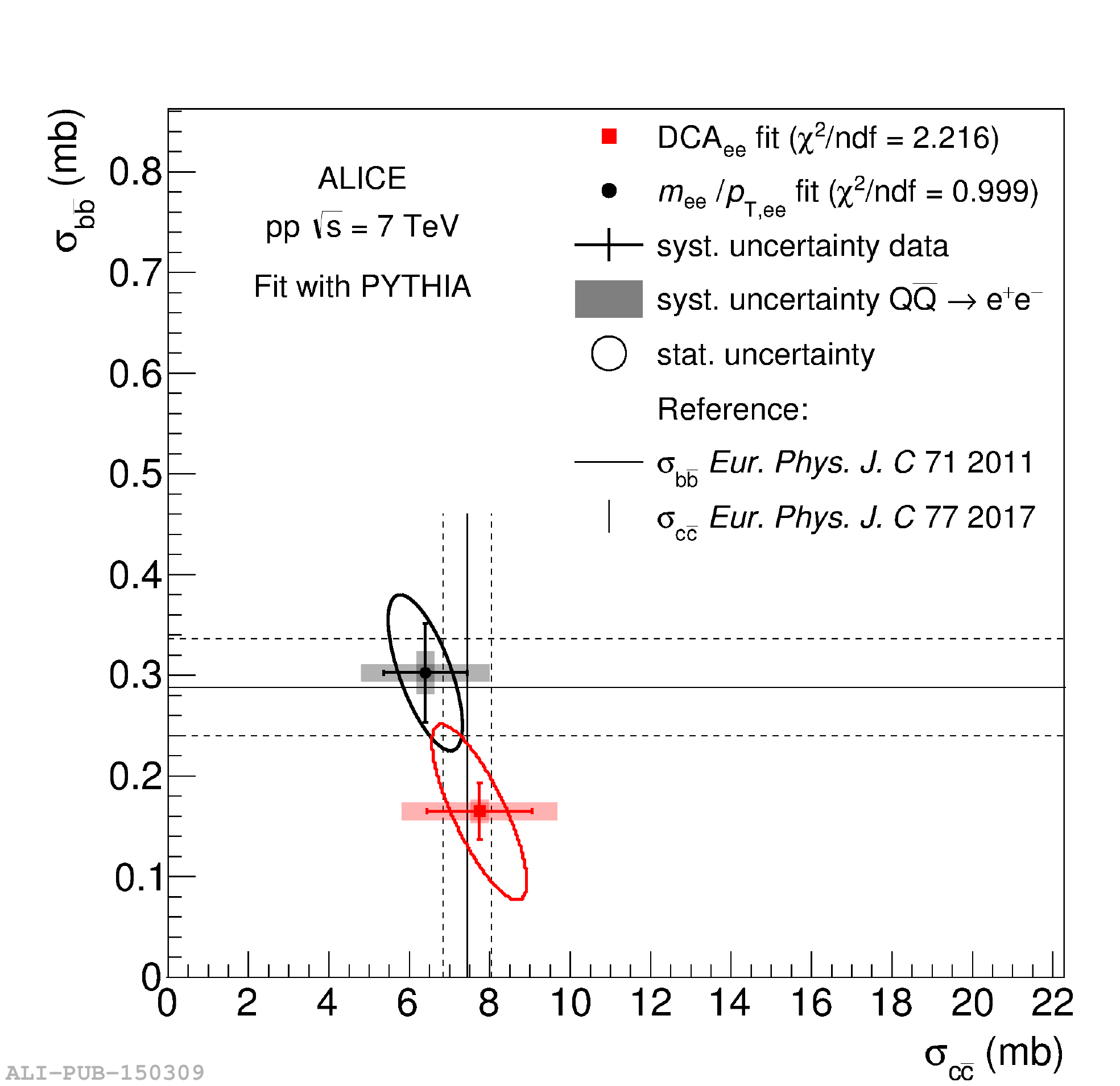}
  \end{minipage}
  \begin{minipage}{0.47\textwidth}
    \includegraphics[trim={0, 0, 0, 1.5cm},clip,scale=0.357]{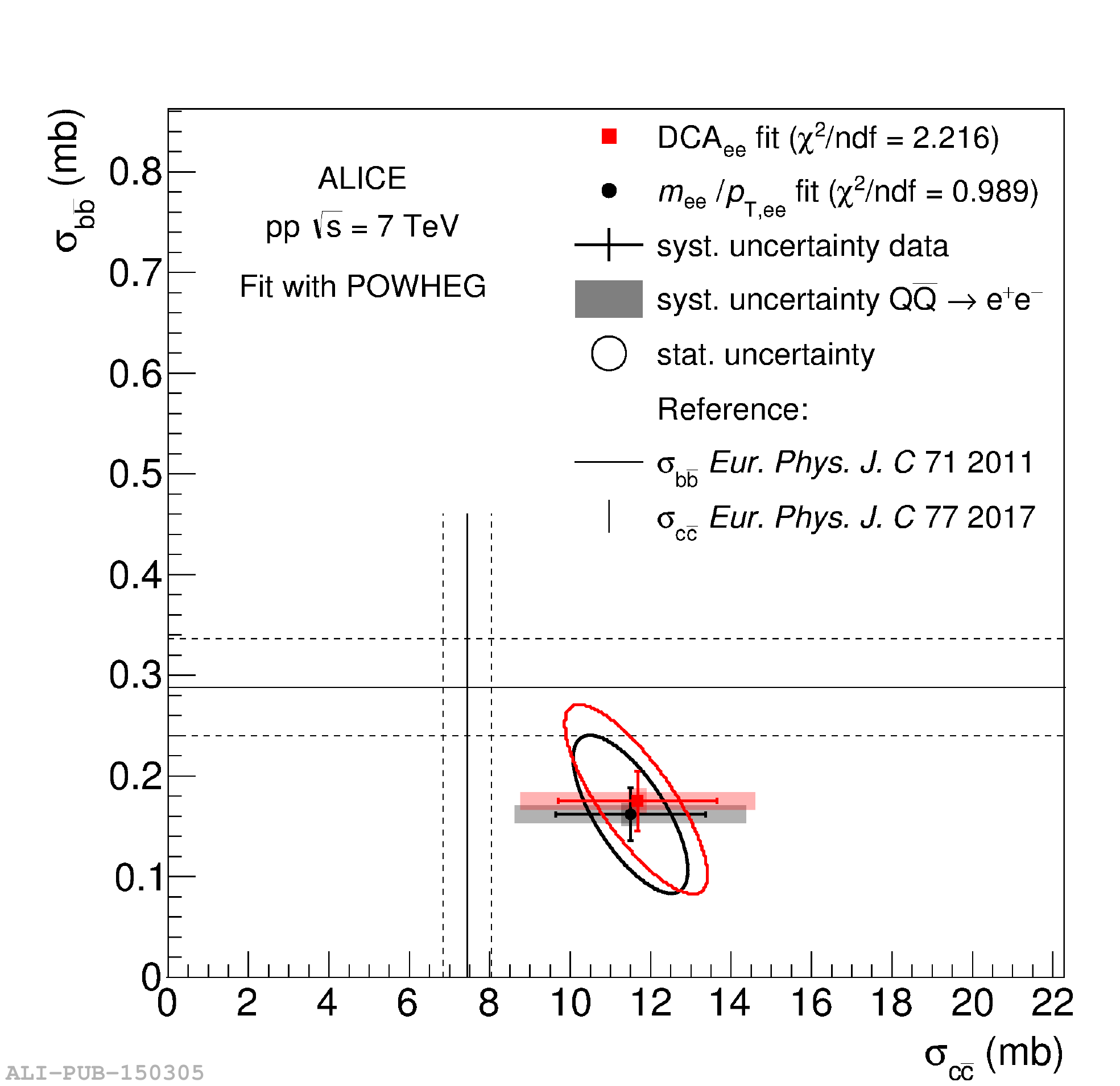}
  \end{minipage}
\caption{Total $\rm c\bar{c}$ and $\rm b\bar{b}$ cross sections with their systematic and statistical uncertainties, extracted from a fit of the measured dielectron yield from heavy-flavour hadron decays in ($m_{\rm ee}$, $p_{\rm T,ee}$) and in \DCAee with PYTHIA (left) and POWHEG (right) are compared to published cross sections (lines)~\cite{ref-ee}.}
\end{figure}
The results are compared to independent measurements of $\sigma_{\rm c\bar{c}}$\cite{ref-ccbar} and $\sigma_{\rm b\bar{b}}$\cite{ref-bbbar} from single heavy-flavour particle spectra and found to be consistent within the large uncertainties. Once these uncertainties are reduced, the dielectron measurements can give further constraints on the MC event generators aiming to reproduce the heavy-flavour production mechanisms.

\section{Conclusion}

To summarise, ALICE measured the dielectron cross sections as a function of $m_{\rm ee}$, $p_{\rm T,ee}$, and \DCAee~in pp collisions at $\sqrt{s} = 7$\,TeV. The hadronic cocktail is in good agreement with the measured  dielectron cross sections in the three discussed observables, which suggests a good understanding of the dielectron cross section in the ALICE acceptance. We show that \DCAee~makes it possible to separate prompt from non-prompt dielectron pairs, and thus will be a key tool to determine the average temperature of the QGP formed in heavy-ion collisions in the future. In the heavy flavour sector we can report a significant dependence of the total cross sections of charm and beauty when using PYTHIA and POWHEG, which reflects the sensitivity of the dielectron measurement to the underlying heavy-flavour production mechanisms implemented in the models.

\bibliographystyle{unsrt}  


\end{document}